# Towards High-Performance Two-Dimensional Black Phosphorus Optoelectronic Devices: the Role of Metal Contacts

Yexin Deng[1, 3*], Nathan J. Conrad [1, 3], Zhe Luo[2, 3], Han Liu[1, 3], Xianfan Xu [2, 3], Peide D. Ye[1, 3 #]

[1] School of Electrical and Computer Engineering, [2] School of Mechanical Engineering and [3]Birck Nanotechnology Center, Purdue University, West Lafayette, IN 47907, USA;   Email: [*]deng58@purdue.edu, [#]yep@purdue.edu

**Abstract**

The metal contacts on 2D black phosphorus field-effect transistor and photodetectors are studied. The metal work functions can significantly impact the Schottky barrier at the metal-semiconductor contact in black phosphorus devices. Higher metal work functions lead to larger output hole currents in p-type transistors, while ambipolar characteristics can be observed with lower work function metals. Photodetectors with record high photoresponsivity (223 mA/W) are demonstrated on black phosphorus through contact-engineering.

## I. Introduction

The discovery of graphene has driven the extensive research interests on 2D materials. However, a lack of bandgap limits its applications on electronic and optoelectronic devices [1-2]. This has led to the recent intensive research on other 2D layered materials with a bandgap, such as $MoS_2$. Recently, black phosphorus (BP) has been found to be an excellent candidate for 2D electronics and optoelectronics due to its high hole mobility (>10000 $cm^2$/Vs) and thickness-dependent direct bandgap [3-6]. BP is a stack of BP monolayers (termed 'phosphorene') with a puckered honeycomb structure, bound together by van der Waals interaction (Fig. 1). Unlike graphene, the bandgap (~0.3 eV or more) of few-layer black phosphorus enables an ON/OFF ratio of >$10^5$ in field-effect transistors (FET), and it has a hole mobility of up to ~1000 $cm^2$/Vs at room temperature [3-5]. The ultra-thin 2D nature makes it promising for use in aggressively scaled FETs, as well as thin film transistors for flexible electronics. Different from $MoS_2$ with a ~1.8 eV direct bandgap only in its monolayer form, BP shows a thickness-dependent direct bandgap, which exhibits ~0.3 eV in its bulk form and increases to >1 eV in its monolayer form -phosphorene [3-4]. Its relatively small direct bandgap makes it ideal for inferred optoelectronic applications. The first BP phototransistor shows a response to wavelength of up to 940 nm. However, it exhibits a relatively low photoresponsivity of 4.8 mA/W [6].

Metal contacts are one of the most important issues for 2D FETs and optoelectronic devices. A clear understanding of metal contacts on BP FETs and photodetectors is imperatively needed to improve the device performance. In this paper, we try to shed a light on the role of various metal contacts on BP FETs and photodetectors. The work function of metal plays an important role on the hole/electron conductions through the Schottky barriers at the metal-BP contacts. FETs with larger work functions metals as contacts exhibit larger hole drain currents, while ambipolar characteristics can be observed on devices with lower work function metals. A photodetector with a record high photoresponsivity (223 mA/V) is demonstrated on BP through contact engineering.

## II. The Role of Metal Contacts on FETs Performance

**a) Device Fabrication**: Few-layer BP flakes were mechanically exfoliated from bulk material onto a p+ doped silicon substrate capped with a 90 nm $SiO_2$. Standard electron beam lithography was used to define the contact patterns. Metal contacts were formed by electron beam evaporation and lift-off process. Fig. 2 shows a schematic of a fabricated BP FET. The $SiO_2$ and p+ Si were used as a gate dielectric, and a back gate, respectively. To study the effects of various metals with different work functions on BP, two different metals were used as contact metals to form two FETs on the same flake (Fig. 3), negating the effect of flake-to-flake variability. Al (4.1 eV), Ti (4.3 eV) and Pd (5.1 eV) were used as metal contacts. The thicknesses of the flakes varied from 5 nm to 20 nm as determined by atomic force microscopy (AFM). The Raman-activated modes in Raman spectra are consistent with the previous work [4], confirming the nature of BP flakes (Fig. 4). All the measurements were performed at room temperature and in ambient atmosphere.

**b) Device Performances and Analyses**: First of all, the electrical characteristics of BP FETs are studied. As shown in the output curves of a typical BP FET using Pd as contacts (Fig. 5), a well-behaved current saturation is observed at a channel length of 1μm. Combined with its ultra-thin channel thickness, which is required for dimensions scaling by Moore's law, it shows the promise of BP for low-power high-speed FET applications. The device on the same flake using Ti as contacts shows a relatively lower on-state current under the same bias conditions. As the device structure is the same, this indicates a larger contact resistance, thus a larger Schottky barrier (SB) at the metal-semiconductor (MS) contact. This phenomenon is observed on all fabricated devices summarized in Fig. 6. Moreover, the devices with Pd as contacts always show larger current no matter how we change the order of metal deposition. The work function of metal is the determinant fact. From the transfer curves of this device (Fig. 7), a larger ON/OFF ratio can be observed on the devices with Ti as contacts. This underlines the nature of BP transistors as SB transistors in this case. As the contact resistance becomes relatively large in Ti device, the modulation of the SB width via back gating allows a larger ON/OFF ratio in Ti device [7]. The switch mechanism on BP FETs is controlled by two SBs as described explicitly in $MoS_2$ FETs [7]. This confirms that the SB height for hole transport in Ti devices is larger than that of Pd devices, and, unlike on $MoS_2$, metal contacts on BP are not strongly pinned. A small ambipolar characteristic can be observed on Ti devices (Fig. 7) and much clearly in Al devices (Fig. 8), as the SB height for electrons in Al is much smaller than that in Ti or in Pd. These results suggest the band diagram of devices with different metals on BP to be like in Fig. 9 (a), in which the SB height for holes at the contact increases as the work function of the metal decreases, and the SB height for electrons decreases as the work function of metal decreases. The strong ambipolar characteristics on Al devices can be illustrated by Fig. 9 (b) and (c). Under negative/positive gate bias, the band bending boosts the hole/electron injection at the MS contact at drain/source as the Al aligns close to the middle of bandgap, resulting in the ambipolar characteristics. Based on these understandings, a Schottky diode with asymmetric contact is demonstrated on the same flake with Ti/Pd as contacts (Fig. 10 and Fig. 11). A rectification ratio of ~100 is obtained and is expected to be larger if using metals with larger work function differences. The conduction mechanism can be understood by the band diagram in Fig. 12. As Pd can make a better p-contact on BP, the Ti SB is mainly used as a Schottky diode.

### III. Contact-Engineered High-Performance Photodetectors

Based on its relatively small and thickness-dependent direct bandgap, BP photodetector is suitable for broadband photodetection. Here we fabricated the photodetectors using the same fabrication process as described above. The photocurrent measurement method is shown in Fig.13. A 633 nm He-Ne laser was focused on the active device region. Based on the study above, the band diagram of the devices with various metal contacts with and without illumination is shown in Fig. 14. Photoresponsivity (R) is defined as $R=I_{ph}/P_{laser}$ where $I_{ph}$ is the photocurrent and $P_{laser}$ is the laser power. With a larger SB height, the photo-generated holes are more easily trapped in the device. This constrains the R of the photodetector, as the photo-generated holes cannot be fully collected. To increase the relatively small R in the first reported BP photodetector [6], here we use Pd as metal contacts in order to reduce the SB height and improve the collection of photo-induced carriers. The device was under a periodic laser illumination and was biased at a small voltage (50 mV). The photoresponse is highly repeatable with long and short illumination period (Fig. 15, 16), demonstrating its capability for photodetection. The detected photocurrent increases as the incident laser power increases (Fig. 17). Moreover, the back gate ($V_{gs}$) can be used to modulate the band diagram, which results in different R with different $V_{gs}$ because the SB changes with $V_{gs}$ or electrostatic doping of the channel (Fig. 18). As the back gate voltage increases, R decreases (Fig. 19). However, the ratio of $I_{illumination}/I_{dark}$, where $I_{illumination}$ and $I_{dark}$ is $I_d$ when the device is with and without illumination, increases as $V_{gs}$ increases, suggests a better signal to noise ratio. These phenomena can be understood by the band diagram in Fig. 20. Under negative $V_{gs}$, the width of the SB near the contact becomes narrower, which makes it easier for holes to go through the barrier. Under positive $V_{gs}$, the width of SB becomes wider which limits the photo-generated electrons limits to reach metal contact and results a lower R. Moreover, the increase of electron density by increasing $V_{gs}$ can also lead to a lower R due to a larger recombination rate in the channel. When $V_{gs}$ is negative, the transistor is turned off, resulting in a larger $I_{illumination}/I_{dark}$ ratio. The maximum R in our devices is 223 mA/V at $V_{gs}$=-30 V and 76 mA/W at $V_{gs}$=0 (Fig. 19), which is 16 times larger than the value reported in [6] at the same $V_{gs}$=0 and with an even smaller $V_{ds}$ in our case. R can be further improved by increasing the $V_{ds}$ (Fig. 21). Finally, Table 1 summarizes the performance metrics of different 2D photodetectors reported in literature for benchmarking.

### IV. Conclusion

1) Metals with larger work function as contacts can significantly reduce the SB height for hole injection on a BP FET to achieve a higher hole drain current. 2) Ambipolar characteristics are observed on the devices using lower work function metals as contacts. 3) High-performance photodetectors have been demonstrated using optimized metal contacts that can reduce the barrier height for photo-generated carrier collection and realize broadband photodetection, showing a record high photoresponsivity (223 mA/W) on this new 2D material.

### Acknowledgement

This work is partly supported by NSF under Grants CMMI-1120577, ECCS-1449270 and ARO W911NF-14-1-0572 monitored by Dr. Joe X. Qiu.

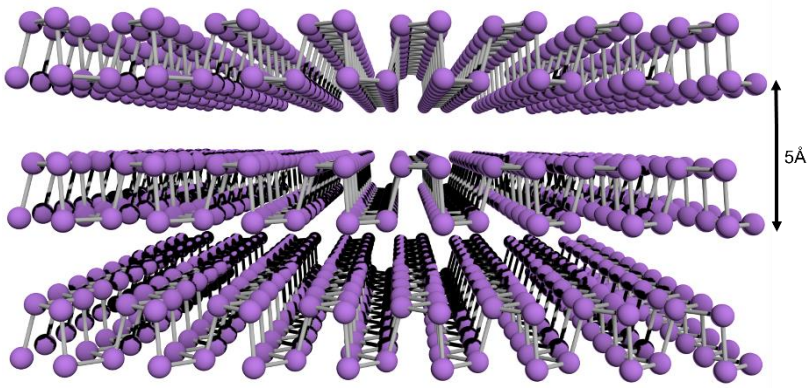

**Fig. 1** Atomic structure of black phosphorus, which is a stack of black phosphorus monolayers or phosphorenes, bound together by van der Waals interactions. In each of the layer, every phosphorus atom is covalently bonded with three adjacent phosphorus atoms to form a puckered honeycomb structure.

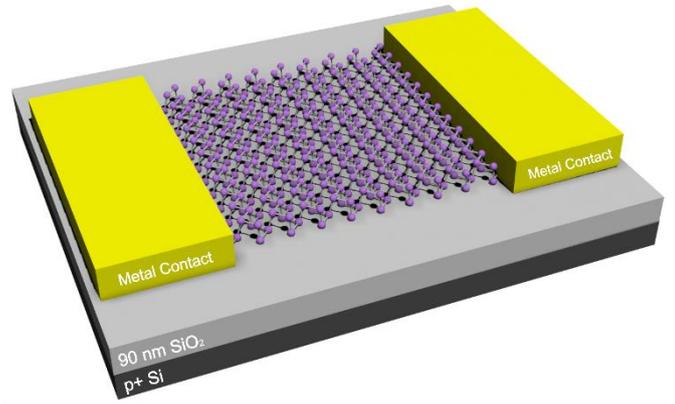

**Fig. 2** Schematic of the black phosphorus field-effect transistor using 90 nm $SiO_2$ as the gate dielectric and p+ doped silicon substrate as the back gate.

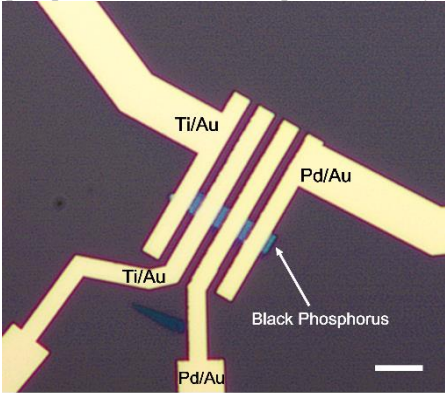

**Fig. 3** Optical image of a fabricated device with different metal contacts on the same black phosphorus flake. Scale bar 5μm.

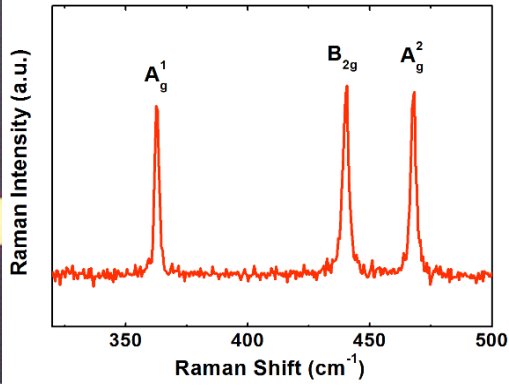

**Fig. 4** Raman spectra of black phosphorus flake. The three Raman-activated modes corresponds to three different vibrational modes

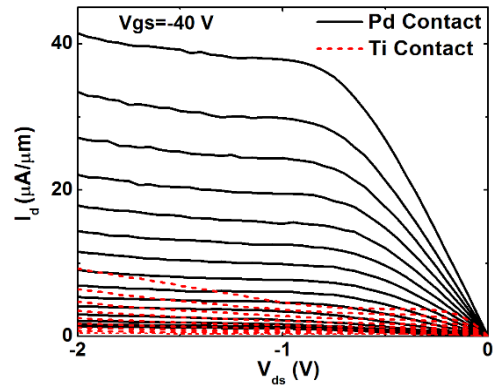

**Fig. 5** Output curves ($V_{gs}$: -40 V-40 V) of two transistors using Pd or Ti as source/drain contacts on the same flake. The Pd device shows larger output currents.

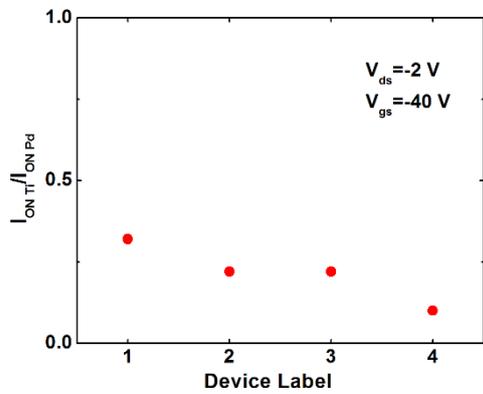

**Fig. 6** The ratio of maximum output current of devices with Pd contacts over Ti contacts. The Pd devices show obviously larger currents.

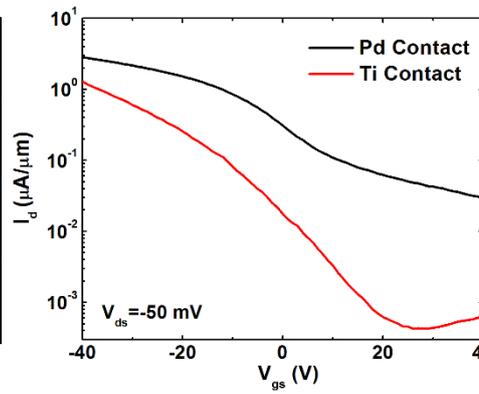

**Fig. 7** Transfer curves of transistors using Pd or Ti as source/drain contacts. The Ti device shows lower current while maintaining larger ON/OFF ratio.

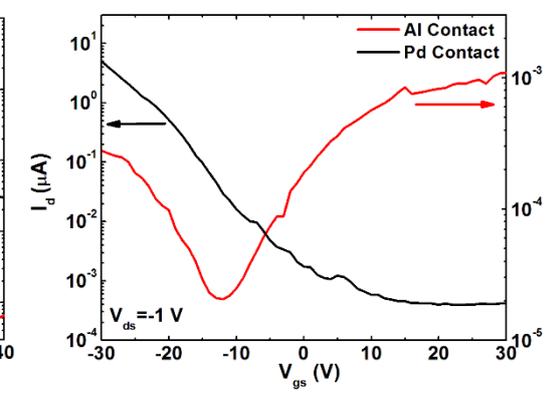

**Fig. 8** Transfer curves of transistors with Pd or Al as source/drain contacts. The device with Al shows obvious ambipolar characteristics.

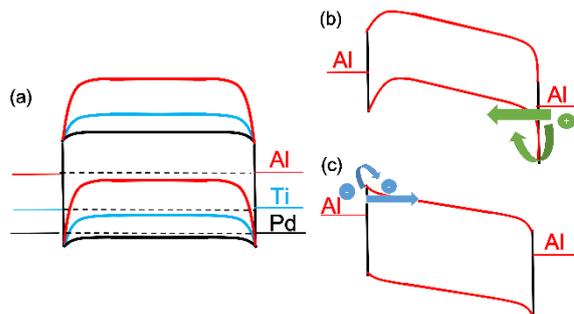

**Fig. 9** (a) Band diagram of a transistor with different metal contacts under zero bias. Band diagram of a device with Al contacts with (b) negative and (c) positive back gate voltages.

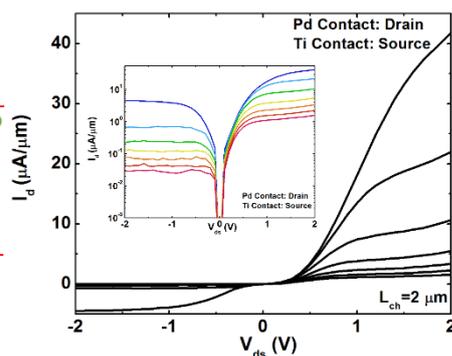

**Fig. 10** Output curves of the device with Ti/Pd as source/drain contacts. The inset shows the curves in semi-log scale.

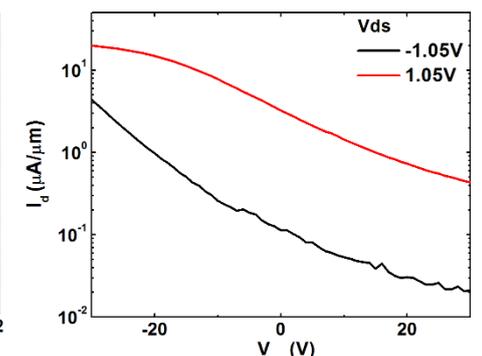

**Fig. 11** Transfer curves of the device with Ti/Pd as source/drain contact with positive and negative $V_{ds}$.

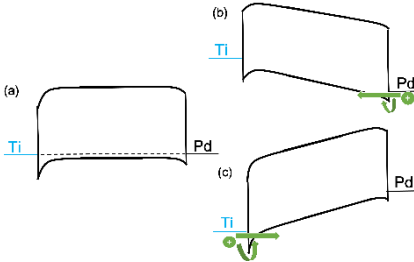
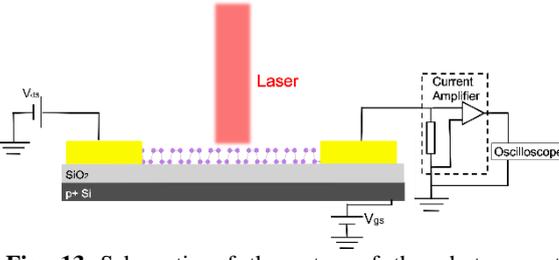
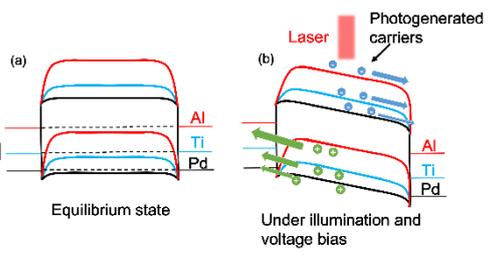

**Fig. 12** Band diagram of the device with Ti/Pd as source/drain contacts (a) equilibrium state, (b) $V_{ds}>0$ and (c) $V_{ds}<0$.

**Fig. 13** Schematic of the setup of the photocurrent measurement under laser illumination. The current in the device is converted to voltage signal, and digitized by an oscilloscope.

**Fig. 14** Band diagram of the device (a) at equilibrium state and (b) under illumination and a small voltage bias ($V_{ds}$) with $V_{gs}=0$.

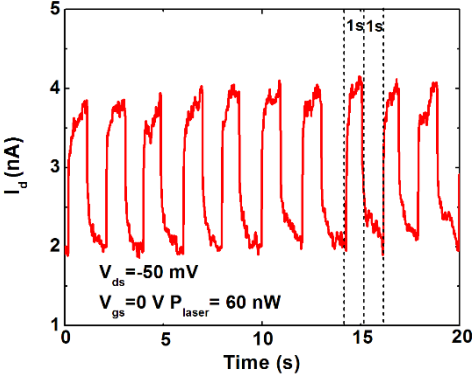
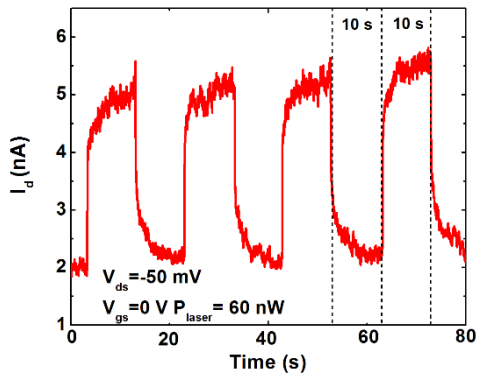
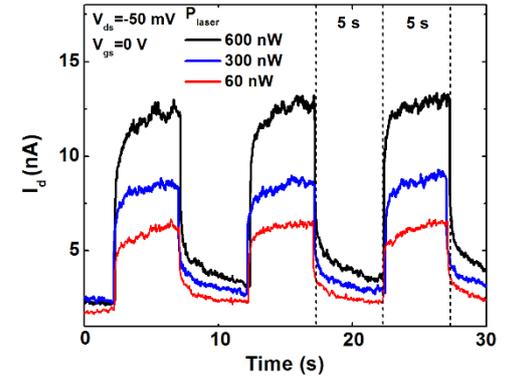

**Fig. 15** $I_d$ as a function of time when the laser is turned on/off with a period of 2 s.

**Fig. 16** $I_d$ as a function of time when the laser is turned on/off with a period of 20 s

**Fig. 17** $I_d$ as a function of time when the laser is turned on/off with a period of 10 s under different incident laser power.

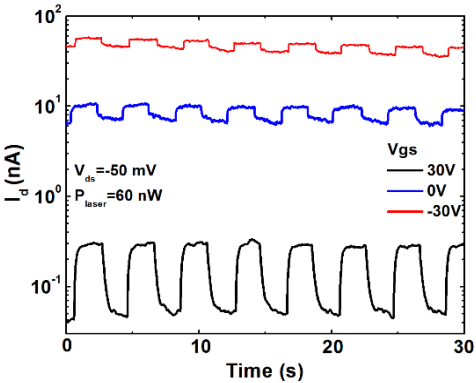
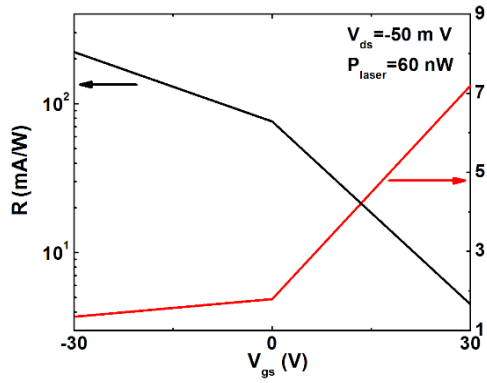
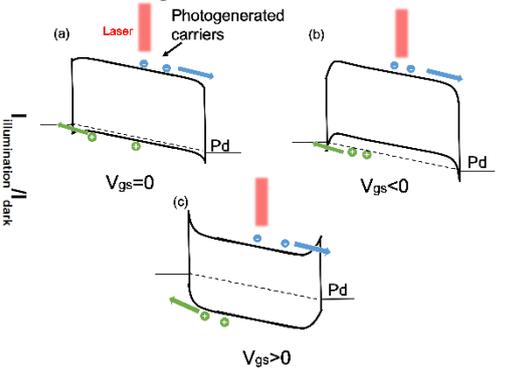

**Fig. 18** $I_d$ as a function of time when the laser is turned on/off with a period of 4 s under different back gate voltages.

**Fig. 19** The photoresponsivity (R) and the $I_{illumination}/I_{dark}$ as a function of back gate voltage. They show different trends when $V_{gs}$ increases.

**Fig. 20** Band diagram of the phototransistor under illumination with (a) $V_{gs}=0$, (b)$V_{gs}<0$ and $V_{gs}>0$.

**Table 1 Summary of 2D Photodetectors**

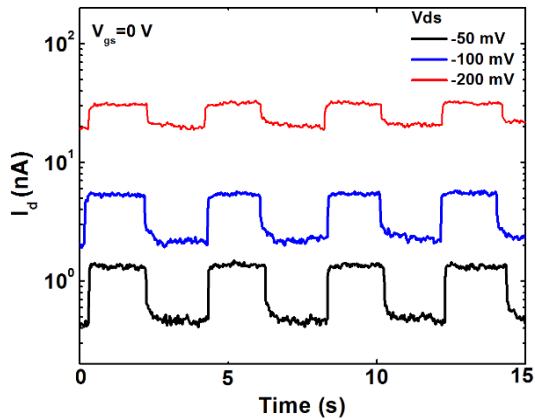

| Materials | Measurement conditions | R(mA/W) | $t_r$ (ms) | Spectral range | Ref |
|---|---|---|---|---|---|
| >1L BP | $V_{ds}$ = 0.05 V, $V_{gs}$ =0V, λ = 633 nm, P = 60 nW | 76 | 100 | Visible−IR | This work |
| >1L BP | $V_{ds}$ = 0.2 V, $V_{gs}$ =0V, λ = 640 nm, P = 10 nW | 4.8 | 1 | Visible−IR | 5 |
| 1L MoS$_2$ | $V_{ds}$ =8V, $V_{gs}$ = −70 V, λ = 561 nm, P = 150 pW | 8.8×10$^6$ | 4000 | Visible | 6 |
| 1L MoS$_2$ | $V_{ds}$ =1V, $V_{gs}$ =50 V, λ = 532 nm, P =80 μW | 8 | 50 | Visible | 7 |
| >1L WS$_2$ | $V_{ds}$ =30 V, $V_{gs}$ = N.A., λ = 458 nm, P = 2 mW | 2.1×10$^{-4}$ | 5.3 | Visible | 8 |
| >1L GaTe | $V_{ds}$ =5V, $V_{gs}$ =0 V, λ = 532 nm, P =3 × 10$^{-5}$ mW/cm$^2$ | 10$^7$ | 6 | Visible | 9 |
| >1L GaS | $V_{ds}$ =2V, $V_{gs}$ =0 V, λ = 254 nm, P = 0.256 mW/cm$^2$ | 4.2×10$^3$ | 30 | UV-Visible | 10 |
| >1L GaSe | $V_{ds}$ =5V, $V_{gs}$ =0 V, λ = 254 nm, P = 1 mW/cm$^2$ | 2.8×10$^3$ | 300 | UV-Visible | 11 |

**Fig. 21** $I_d$ as a function of time when the laser is turned on/off with a period of 4 s under different $V_{ds}$.